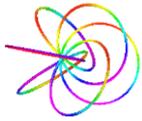

# CONDUCT RISK: DISTRIBUTION MODELS WITH VERY THIN TAILS

Peter Mitic[1, 2, 3]

1: Santander UK
2, Triton Square, Regent's Place, London NW1 3AN
e-mail: peter.mitic@santandergcb.com
2: Department of Computer Science, University College London
Gower Street, London WC1E 6BT
3: LabEx Refi

## Abstract

*Regulatory requirements dictate that financial institutions must calculate risk capital (funds that must be retained to cover future losses) at least annually. Procedures for doing this have been well-established for many years, but recent developments in the treatment of conduct risk (the risk of loss due to the relationship between a financial institution and its customers) have cast doubt on 'standard' procedures. Regulations require that operational risk losses should be aggregated by originating event. The effect is that a large number of small and medium-sized losses are aggregated into a small number of very large losses, such that a risk capital calculation produces a hugely inflated result. To solve this problem, a novel distribution based on a probability density with an $\exp(-x^4/(2s^2))$ component is proposed, where s is a parameter to be estimated. Symbolic computation is used to derive the necessary analytical expressions with which to formulate the problem, and is followed by numeric calculations in R. Goodness-of-fit and parameter estimation are both determined by using a novel method developed specifically for use with probability distribution functions. The results compare favourably with an existing model that used a LogGamma Mixture density, for which it was necessary to limit the frequency and severity of the losses. No such limits were needed using the $\exp(-x^4/2)$ density.*

## Disclaimer

# 1. INTRODUCTION

Regulated financial institutions (hereinafter referred to as 'banks') are required annually to assess the amount of capital that must be retained to cover operational risk losses that might be suffered in the following year. The European Banking Authority (EBA) defines *Operational Risk* as "*the risk of losses stemming from inadequate or failed internal processes, people and systems or from external events. Operational risk includes legal risks but excludes reputational risk and is embedded in all banking products and activities.*" [1]. Informally, operational risk is the risk of "things going wrong". Such capital must be retained by the bank and cannot be used for lending. The amount retained should be enough to cover expected losses, which are typically a mixture of known liabilities and averages based on known losses for prior years. In addition, the amount retained should also include an estimate to cover 'unexpected' losses that cannot be anticipated as individual amounts. The amounts actually retained vary greatly depending on the size of the bank: from a few hundreds of millions of euros for a small retail bank to many billions for a large international investment bank. The requirement to retain funds, but also and to make funds available for lending and investment, always creates conflict: the two activities are contradictory. A balance must therefore be struck when calculating the amount of capital to retain. It should be enough to satisfy regulations, but should not so excessive as to hinder business activity. The calculation of the capital amount is therefore an important part of a bank's risk control function.

The Basel Committee on Banking Supervision (often known as the "Basel Committee") is the originating organisation for regulations that govern the management of operational risk, and for regulations on the calculation of operational risk capital [2]. At the time of writing three principle streams for calculating of operational risk capital are in operation. The *Basic Indicator Approach* is based on annual revenue, whereas the *Standardized Approach* uses annual revenue of business lines. The third stream is the Advanced Measurement Approach (AMA), in which each bank is permitted to develop its own risk model, provided they are consistent with the regulations in [2]. The model proposed in this paper falls into the AMA category. Those regulations give a very strong indication of what has to be done, and what might be described as a "standard model" has emerged. Such a "standard model" typically has the following components:

- A statistical model of severity (i.e. magnitude) of losses, based on a 'fat-tailed' distribution (i.e. one for which the probability that a very large loss occurs is relatively large – a precise definition will be given in the Methodology section)
- A statistical model of loss frequency: often Poisson
- A convolution model to combine the severity and frequency models, and extract the 99.9% value-at-risk (VaR). VaR can be thought of as 99.9 percentile loss, the figure being specified by the Basel regulations [2]. The precise method is given in the Methodology section.

Conduct Risk is a significant component of operational risk. Some banks treat it as a separate entity, but a capital calculation for it is needed whatever its classification. The Basel regulations define a risk class taxonomy, and conduct risk is part of the 4th category in Table 1 below.

| Basel ID | Basel Category | Abbreviation | Components |
|---|---|---|---|
| 1 | Internal Fraud | IF | misappropriation of assets, tax evasion, bribery |
| 2 | External Fraud | EF | hacking, third-party theft, forgery |
| 3 | Employment Practices and Workplace Safety | EPWS | discrimination, compensation, employee health and safety |
| 4 | Clients, Products, and Business Practice | CPBP | market manipulation, antitrust, improper trade, product defects, legal actions |
| 5 | Damage to Physical Assets | DPA | natural disasters, terrorism, vandalism |
| 6 | Business Disruption and Systems Failures | BDSF | disruptions, software and hardware failures |
| 7 | Execution, Delivery, and Process Management | EDPM | data entry errors, accounting errors, failed mandatory reporting, negligent loss |

Table 1. Basel Risk Class taxonomy

Section 2 gives more details of the nature of conduct risk.

## 2. CONDUCT RISK

"Conduct Risk" (CR) may be regarded as "risk that arises as a result of how firms and employees conduct themselves, particularly in relation to clients and competitors" [3]. Its essential element is how a bank behaves with respect to its customers and other stakeholders, and encompasses items such as compensation resulting from complaints, regulatory fines and costs of mis-selling. Mis-selling of Payment Protection Insurance (PPI) is a significant part of CR losses in the UK. They, in particular, are the losses that drive the proposed model in this paper.

## 2.1. The Source of Conduct Risk Losses

Thomson Reuters [4] gives a somewhat alarming report on how a selection of firms view CR: "*81 percent of firms remain unclear about what conduct risk is and how to deal with it*." This report gives no indication of the sample size, but it does signal that problems exist. It does not say that there is a general problem in how losses should be classified within the taxonomy in Table 1. It is not always clear which category a particular loss should be allocated to. Such classification problems are mostly invisible to the statistical modeller, who sees the end result: a sequence of numbers with a timestamp and a Basel risk category for each. Thomson Reuters [5] lists, according to their respondents, the principal sources of conduct risk. In terms of frequency of response, the most significant five are (in decreasing order of frequency):

1. culture;
2. corporate governance;
3. conflicts of interest;
4. reputation;
5. sales practices.

Of these, all except "reputation" enter into the modelling process. The remaining categories map to the Basel sub-category "improper trade" in Table 1. Payment Protection Insurance is easy to identify as belonging to the CPBP category. It is a particular problem in the UK, where customer compensations have been huge. The Financial Conduct Authority (FCA) defines: "*Payment Protection Insurance (PPI) is designed to cover your debt repayments in circumstances where you aren't able to make them such as accident, redundancy, or illness.*" [6]. It was found that customers were sold PPI in conjunction with other financial products (loans, mortgages etc.) without telling them that PPI was an optional extra, and without explaining the product adequately. The Guardian [7] reports massive PPI payouts in Figure 2. Note that the amounts in the graphic sum to £39.7bn, to which £600m should be added (the latest payout at the time from Barclays Bank and the subject of the article, making a total of £40.3bn.

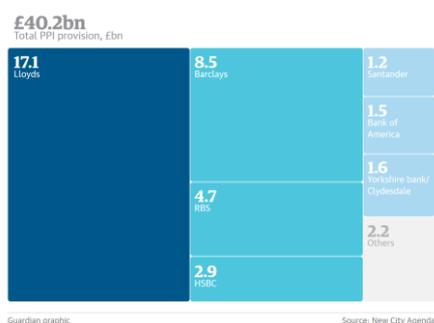

Figure 1. PPI payouts to October 2016

## 2.2. Characteristics of Conduct Risk Losses

Conduct risk losses comprise, typically, thousands of small- to medium-sized-payments (ranging from tens of pounds to a few thousands) made to individual customers. In addition there will be costs such as legal fees and regulatory fines. Both of those could be large: of the order of millions of pounds. As such, 'fat-tailed' distributions serve very well in capital models. They can model precisely the situation that exists: a very large number of small losses totalling a small proportion of the total loss, with a small number of large losses making up the balance of the total. It is not unknown for the largest 5% of losses to be worth 95% of the total loss. Compared with other categories of operational risk losses, conduct risk losses in the form of customer payments are an extreme. Their annual frequency is higher, their mean loss is smaller, and the calculated capital tends to be larger due to the high frequency.

The Basel regulations [2] direct that operational risk losses should be aggregated by originating event. Such aggregation is apparently inconsistent with using individual payments to customers. Originating event is not the only way to aggregate, although others may not be Basel-compliant. Consider the case where a decision to market a product is distributed regionally throughout a firm, and that the distribution is rolled out over an extended period covering several years. Each individual action within the roll-out scheme could be considered a root event for all the losses arising from it. The losses are therefore grouped by time and by region. The rollout could be further subdivided by branch, sales force or customer category. That adds another layer of potential root events. These simple considerations show that the term *root event* is not well defined. There is a fuller discussion of this topic in [8].

Against the argument in favour of aggregation is the view that an originating event should be the contract made between a bank and each individual customer. Without such a contract, there would be no breach of contract and no compensation payment due to inappropriate conduct. In the absence of any specific regulatory directive, most banks take the view that conduct risk losses should be aggregated, either by originating event, or by accounting period.

The result of aggregation of conduct risk losses is a very small number of very large 'losses', since common practice is to define very few originating events. Hundreds of thousands of small losses are thereby condensed into only ten, twenty or thirty 'losses'. This causes considerable capital modelling problems because existing 'fat-tailed' capital models do not cope well with a small number of large losses. Effectively the small number of large losses comprises the tail of a distribution (i.e. very high losses that occur with a very small probability), and 'fat-tailed' models estimate values in the tail of a tail, which are "super high"! They are effectively outliers of a set that does not exist.

In principle, conduct risk may be modelled in exactly the same way as any other operational risk, provided that there are not too few losses to find a reasonable distribution fit. The result of calculating capital using only the tail of a distribution is a grossly inflated value. To give an example, we recently modelled a set of about 190000 conduct risk payments which had been aggregated into 30 aggregated losses. The capital calculated from the payments data was in the order of £260m, but the capital calculated from the aggregations was near to £3700m. If we summarise the aggregation process, we could say that the aggregation takes us from one extreme to the other: a large number of small losses to a small number of large losses.

## 3. MODELLING METHODOLOGY

This section describes the basis of the standard operational risk capital calculation. The method depends on modelling loss severity by an appropriate probability distribution. The advance presented in this paper is to find a probability distribution that models the severity of conduct risk losses in a more satisfactory way than 'fat-tailed' distributions. In all cases, regulatory capital is assessed in the same way, as described in the next section.

### 3.1. Data

This section illustrates some examples of aggregated conduct risk losses. Figure 2 shows three distinct examples. The histograms show single instances of very large losses covering a five year period. Compared with other operational risk losses, all are outliers in the sense that all are atypically large.

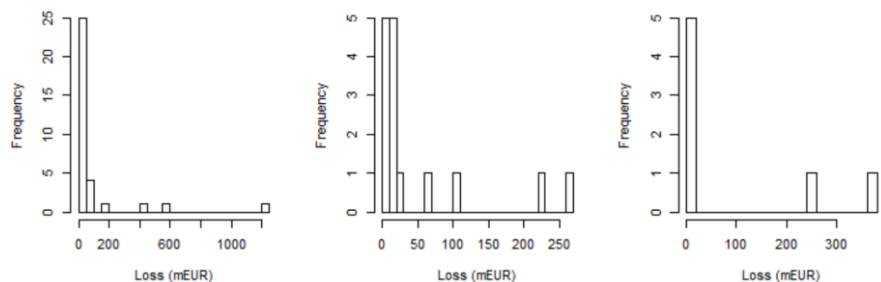
Figure 2. Typical conduct risk loss histograms

The minimum, maximum and mean losses (in mEUR) for the examples in Figure 2 are:

| | |
|---|---|
| Left-hand data set: | (0.1, 1208.0, 92.7) |
| Middle data set: | (1.0, 264.2, 52.5) |
| Right-hand data set: | (0.6 362.0, 91.7) |

Using these and similar data sets causes difficulties for determining parameters of a severity distribution. Iterative methods, such as *maximum likelihood*, do not always converge, and if they do, they sometimes converge to values that are clearly wrong. Usually such a failure is due to a flat feature space. Gradient search methods produce essentially random results because the result of each iteration is very similar to the result of the previous iteration.

## 3.2. The Capital Calculation

The capital calculation assumes that a suitable severity and a suitable frequency distribution has been fitted to empirical data. Any such pair of distributions should have passed an appropriate goodness-of-fit (GoF) test. The method described in this section is due to Frachot et al [9], and is known as the Loss Distribution Approach (LDA). It is a Monte Carlo process based on a convolution of the severity and frequency distributions. Other methods are available, notably Panjer recursion and the Fast Fourier Transform (FFT). See [10] for details of all three, and also of how they are used in the general context of operational risk.

*Algorithm LDA* is a summary of the LDA method. For a set of *N* losses covering a period *y* years, and let *n* be the number of Monte Carlo trials used. Further, let the fitted severity distribution to those losses be $F(l, \boldsymbol{p})$, where *l* is a loss and $\boldsymbol{p}$ is a vector of parameters obtained from the fit.

*Algorithm LDA*

a) Calculate an annual loss frequency, $\lambda = N/y$
b) Repeat *n* times
   i) Obtain a sample size *z* by drawing a random sample of size 1 from a Poisson($\lambda$) distribution (*)
   ii) Draw a sample of size *z*, $S_z$, from the severity distribution, *F*.
   iii) Sum the losses in $S_z$ to obtain $\sum S_z$, the estimate of annual loss.
c) End Repeat
d) Calculate the 99.9$^{th}$ percentile of the retained sums $\sum S_z$ (99.9 is the percentage specified by the Basel regulations [2].

(*) A Poisson distribution is not the only way to do this. A negative binomial distribution is often used. For large $\lambda$, a normal approximation to the binomial is appropriate.

*Algorithm LDA* is easy to implement and is applicable for any severity distribution. In practice, *n* might exceed one million before stability of the result is established. In that case, the Monte Carlo process might take one or two hours. In most cases 100000 iterations suffice and the calculation takes minutes.

### 3.3. Fat-tailed distributions

An account of "fat-tailed" distributions may be found in [11, 12]. Equation (1) gives a characterisation for the density function *f(x)* and the corresponding distribution function *F(x)* for a random variable in terms of a parameter *a*. For large *x*, these functions are polynomial-like rather than exponential-like.

$$\left.\begin{array}{ll} f(x) \sim x^{-(1+a)} & a > 0; x \to \infty \\ F(x) \sim x^{-a} & a > 0; x \to \infty \end{array}\right\} \quad (1)$$

Typical examples of "fat-tailed" distributions are the LogNormal, with distribution function (in terms of the Normal distribution function $\Phi$) $F(x) = \Phi\big((\ln(x) - \mu)/\sigma\big); \mu \in \mathbb{R}, \sigma > 0$, and the Weibull, with distribution function $F(x) = 1 - e^{\left(\frac{x}{\theta}\right)^{\tau}}; 0 < \tau < 1, \theta > 0$. Normalisation factor for density. The Pareto distribution, $F(x) = 1 - \left(1 + \frac{x}{\theta}\right)^{-\alpha}; \theta, \alpha > 0$ is particularly troublesome because it nearly always returns very high 99.9% VaR values.

The 99.9% VaR values obtained using "fat-tailed" distributions with data as illustrated in section 3.1 are always intuitively unacceptably high, which prompts a search for an alternative. In many cases the capital value is orders of magnitude greater than the expected value. *Algorithm LDA* works well using a "fat-tailed" distribution for data where there is a mixture of small- to mid-value losses with some very large losses. Aggregated losses effectively constitute the tail of a distribution that has a missing body.

### 3.4. Thin- and Very-thin-tailed distributions

In order to use *Algorithm LDA* with aggregated conduct risk losses, the required distribution is exactly the opposite of a "fat-tailed" distribution: namely a "thin-tailed" distribution. The distinguishing characteristic of such a distribution is that the probability of generating an extreme value in a random sample should be much less than the probability of generating an extreme value using an exponential-based distribution (the "low loss = high probability; high loss = low probability" criterion). A first approximation to a "thin-tailed" distribution is the Normal distribution. This proved to be easy to do but still resulted in a capital value which was unreasonably large. An alternative is distribution which resembles the Normal distribution but was capable of generating relatively fewer very high value losses in a random sample. In parallel there should also be a relatively high probability of generating lower valued losses. Such a distribution can be obtained, by replacing the $<< e^{-\frac{x^2}{2}} >>$ term in the standard Normal density by $<< e^{-\frac{x^4}{2}} >>$, and by choosing a suitable domain for *x*.

The density plot for such a distribution resembles that of the Normal distribution, but decays much faster for losses that are a large distance from the mean loss. This is a big advantage but the disadvantage is that the probability of generating very small losses in a random sample is small. The actual density proposed introduces a scale parameter *s*, to be determined from the data, so that the non-normalised density contains $<< e^{-\frac{x^4}{2s^4}} >>$. The resulting density is termed the *Exp4* distribution, and its formal probability density function *f(x,s)* is given in equation (2).

$$f(x,s) = \frac{1}{s2^{\frac{1}{4}}\Gamma\left(\frac{5}{4}\right)} e^{-\frac{x^4}{2s^4}}; \quad s > 0, x > 0 \tag{2}$$

The normalising factor $s2^{\frac{1}{4}}\Gamma\left(\frac{5}{4}\right)$ is produced by Mathematica, and the final expression for *f(x,s)* requires inclusion of appropriate assumptions to make the result useable. Appendix A shows the details, and an illustrative graph. The density definition shown is the most convenient way to formulate *f(x,s)*. The fourth power in the exponential term ensures that the probability of generating very high values in a random sample is low, and the domain *x*>0 ensures that the probability of generating low values in a random sample is high (as the density graph in Appendix A shows).

The corresponding cumulative distribution function, *F(x,s)* is given in equation (3).

$$F(x,s) = \int_0^x f(t,s)dt = 1 - \frac{1}{\Gamma\left(\frac{1}{4}\right)}\Gamma\left(\frac{1}{4}, \frac{x^4}{2s^4}\right); \quad s > 0, x > 0 \tag{3}$$

The integral of *f(x,s)* in equation (3) is again provided by Mathematica, which requires explicit assumptions $s > 0$ and $x > 0$ to avoid generating an unwieldy expression involving an exponential integral (function ExpIntegral[] in Mathematica) See Appendix A for the appropriate expressions. In equation (3), the two-parameter form of the Gamma function is the *upper incomplete gamma* function, $\Gamma(a,x) = \int_x^\infty t^{a-1} e^t dt$. See [13] for details. The shape of the *Exp4* density and distribution functions is unlike those of "fat-tailed" distributions. Most notably, "fat-tailed" distributions typically have near vertical slopes for small losses. The *Exp4* distribution is set up to use scaled losses: specifically each loss divided by the mean loss. This scaling ensures that the arguments *x* in (2) and (3) are small positive real numbers which are consistent with the domains indicated.

In the R statistical language, which was used for the numerical calculations in this paper, the *upper incomplete gamma* function is implemented in the *gsl* package by the *gamma_inc*() function. It is customary in R to define four functions per probability distribution: one for the density, one for the distribution, one for the inverse distribution, and one for random number generation. The names of these functions are, by convention, prefixed by *d, p, q* and *r* respectively. The R implementation for the *Exp4* set is given in Appendix B.

## 3.5. Parameter Estimation and Goodness-of-fit

Given the *Exp4* distribution and data as described in the previous sections, the usual way to proceed is to estimate the distribution parameters, and use them in a GoF test to assess the quality of the fit. Attempting fit the *Exp4* distribution to aggregated loss data using standard maximum likelihood methods proved to be difficult in R. Parameter values that were clearly incorrect were often returned. This was attributed to using a gradient search in a parameter space for which even large changes in parameter values had little effect on an objective function. This situation can be likened to an attempt to climb hills in an almost flat landscape. A search for an optimal solution proceeds in an almost random direction, and any 'optimal' solution found is only locally optimal.

As an alternative, parameter optimisation was combined with a GoF test using the author's *TN* ("Transformed Normal") method [14]. There are three *TN* tests, and the simplest conceptually and in practical terms is the *TN-A* test. All were originally intended as dedicated GoF tests for cumulative distribution functions obtained from fitting "fat-tailed" distributions to operational risk data. The *TN-A* test has the advantages the following advantages over 'traditional' GoF tests such as Anderson-Darling (AD) or Kolmogorov-Smirnov (KS).

- It satisfies the criterion "*if a fit looks good, the test should say so*". The AD and KS tests were found to reject distributions that that were intuitive good fits, especially for large data sets.
- It is independent of the number of data points.
- It is entirely deterministic.
- The value of the *TN-A* statistic (i.e. the test's objective function) is a direct measure of the goodness-of-fit. The smaller then *TN-A* value, the better the fit. In contrast, the AD and KS statistic values only indicate whether or not a null hypothesis should be rejected.

The way in which the *TN-A* statistic is formulated implies that it can also be used a data fitting methodology. That is the approach used to fit the *Exp4* distribution to conduct risk loss data, and is the first time the *TN* formulation has been used in this way. The following section therefore explains the basic concepts of the *TN* formulation.

### 3.5.1. The *TN-A* method: goodness-of-fit

Given a set of *n* real numbers $X = \{x_1, x_2, \ldots, x_n\}$ (which are the aggregated conduct risk losses), assign a probability $y_i$ to each loss $x_i$, where $y_i = (i - 0.5)/n$. The probability $y_i$ is this a cumulative probability. The set $D = \{x_i, y_i\}_{i=1}^{n}$ then defines the cumulative empirical distribution of X. the requirement is to fit a probability distribution to D.

Suppose that, in general, a fit is proposed using a distribution function $F(x, \boldsymbol{p})$, where $\boldsymbol{p}$ is a vector of parameters *that has already been* determined. If the fit is reasonable, a plot of the set D superimposed on a plot of $F(x, \boldsymbol{p})$ should resemble the left-hand part of Figure 3. Notice the relative positions of the points of D to the curve defined by $F(x, \boldsymbol{p})$. There are consecutive chains of points that are either below or above the $F(x, \boldsymbol{p})$ curve. If the points in these chains are joined by straight lines, very few of those joining lines cross the $F(x, \boldsymbol{p})$ curve. That is an accurate assessment of what happens with actual data. The combination $\boldsymbol{L} = \{D, F(x, \boldsymbol{p})\}$ was referred to as *Loss-Space* in [14], because actual losses are involved. The right-hand part of Figure 3 will be explained below. It is obtained by applying a transformation T to the points in $\boldsymbol{L}$, thereby deriving *Probability-Space* (so-called because the transformation is based on a probability measure.

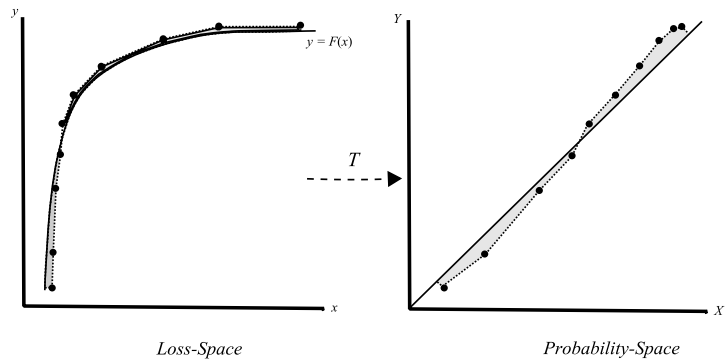

Figure 3. *Loss-Space* transformed to *Probability-Space*

Consider now the effect of the mapping T, defined in equation (4), below.

$$\left. \begin{array}{l} T: \{\mathbb{R}^+ \otimes (0,1)\} \to (0,1)^2 \\ T(x, y) \to (F(x, \boldsymbol{p}), y) = (X, Y) \end{array} \right\} \quad (4)$$

Under T, the elements $y_i$ in D are unchanged, other than to be renamed $Y_i$. The elements $x_i$ in D are mapped by the distribution function, and are labelled $X_i$. The result is that *Probability-Space* is the unit square. the $F(x, \boldsymbol{p})$ curve always maps to the line $Y = X$. This makes it much easier to analyse the "Enclosed Area" in *Probability-Space*, which is the area enclosed between the line $Y = X$ and the mapped points $T(x_i, y_i)$ once it is joined by straight line segments. This is the shaded area in the right-hand part of Figure 3. Calculating that "Enclosed Area" in *Probability-Space* is very easy. The smaller it is, the better the fit. That is the essence of the *TN* group of GoF tests.

The significance level for a data fit using the *TN-A* test can be calculated very easily. An elaborate topological argument in [14] leads eventually to a simple quadratic expression for the significance level (equation 5). In (5), *p* is a probability and *A(p)* is the "Enclosed Area".

$$A(p) = 2\sqrt{2p}(1 - \sqrt{2p}) \qquad (5)$$

To obtain a *p*% 2-tailed significance level for the *TN-A* statistic, the value *p*/2 (expressed as a probability rather than a percentage) should be used in (5). For example, the 5% significance level is calculated using $p = 5/(2 \times 100) = 0.025$, for which $A(0.025) = 0.0682$. If the calculated "Enclosed Area", $\tilde{A}$, is less than 0.0682, the fit "passes the significance test at 5%". A more rigourous restatement of that phrase is given in Appendix C.

### 3.5.2. The *TN-A* method: parameter estimation

Focussing now on the best fit itself, by supplying values for the *Exp4* distribution parameter *s*, an array of *TN-A* values can be built, and the minimum selected. The value of *s* corresponding to the minimum *TN-A*, $\hat{s}$, is then the required parameter value. In practice $\hat{s}$ is found by starting with an initial estimate, followed by an ordered search. In practice the ordered search is likely to be embedded in a function call. In R this is the function `optimize()`, and its equivalent in Mathematica is `FindMinimum[]`.

Note, however that using the *TN-A* test to provide a best fit only works well in the current context because only one parameter needs to be estimated. Possibly, searching a bivariate parameter space would also work reasonably well, but would take longer. To search a parameter space with more than two parameters would probably require a sophisticated algorithm in order to search efficiently.

Using the *TN-A* method for parameter estimation as well as GoF automatically produces an optimal fit with an assessment of how good that fit is.

### 3.6. Extension to higher powers of *x/s*

Use of a density containing an $<< x^4 >>$ term prompts the question of whether or not a higher power of *x* would be appropriate. One has to be careful not to over-fit in these circumstances. It may be possible to find a distribution that either fits the data better or produces a lower capital value, but that distribution type may not work so well with other data sets. To see the effect of an extension if the *Exp4* distribution to higher powers of *x*, the same procedures as were used in section 3.4 were applied to higher even powers of *x*.

Equations (6) and (7) define the density *fn*() and distribution *Fn*() functions respectively for the required generalisation: the *ExpN* distribution. As with the *Exp4* distribution, Mathematica was used to derive them.

$$fn(x,s) = \frac{1}{s 2^{\frac{1}{n}} \Gamma\left(1+\frac{1}{n}\right)} e^{-\frac{x^n}{2s^n}}; \quad s > 0, x > 0, n \epsilon \mathbb{Z}^+ \tag{6}$$

$$Fn(x,s,n) = 1 - \frac{1}{\Gamma\left(\frac{1}{n}\right)} \Gamma\left(\frac{1}{n}, \frac{x^n}{2s^n}\right); \quad s > 0, x > 0, n \epsilon \mathbb{Z}^+ \tag{7}$$

A quick check shows that when *n* = 4, *ExpN* reduces to *Exp4*. A useful test of the applicability of the *ExpN* distribution is to consider the 99.9 percentile for selected values of *n*. Figure 4 shows a comparison of these percentiles for even values of *n* between 6 and 20, together with the corresponding *Exp4* percentile. It is clear from Figure 4 that there is a reduction in the 99.9 percentile value as *n* increases, but that the rate of decreases diminishes. Very little further decrease is observed for *n* ≥12. Therefore the case *n* = 12, is a potential rival to the *Exp4* distribution, but there is a warning against using it in Section 5.

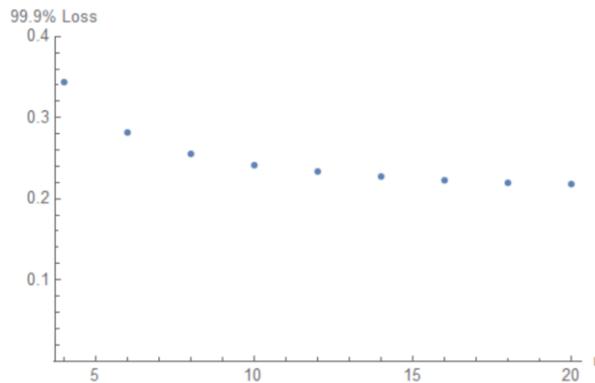

Figure 4. Variation of *ExpN* 99.9 percentiles with *n*

## 4. RESULTS

Capital values for each of twelve sets of aggregated conduct risk losses were calculated using Algorithm LDA and the *Exp4* distribution. Histograms of three of them are shown in Figure 2, section 3.1. All twelve have the same characteristics of having only a few large losses, with no small or medium sized losses. Simple statistics (number and sum of losses) are indicated in Table 2, below. Table 2 also shows the capital values (99.9% VaR) for each of them with their *TNA* test values. In the previous section the *ExpN* distribution with parameter value *n* = 12 was introduced as a potential alternative to *Exp4*. Instead of restricting *n* to a modestly low value, consider a much higher value, *n* = 100, as a proxy for infinity.

Therefore, as a comparison, the Table 2 also shows capital values obtained by fitting a Normal distribution and the *ExpN* distribution with $n = 100$ (i.e $n \to \infty$) to each data set. The sum and capitals shown are in millions of euros.

| Data Set | Count | Sum | Capital | TNA | Normal Capital | *ExpN* ($n \to \infty$) Capital |
|---|---|---|---|---|---|---|
| 1 | 29 | 1876.8 | 539.2 | 0.110 | 1825.7 | 516.2 |
| 2 | 21 | 2110.9 | 395.3 | 0.074 | 2673.4 | 365.7 |
| 3 | 33 | 3059.4 | 503.8 | 0.098 | 3045.6 | 471.2 |
| 4 | 17 | 2287.2 | 1670 | 0.031 | 1728.5 | 1557.1 |
| 5 | 29 | 3437.9 | 1390.1 | 0.036 | 2411.5 | 1295.3 |
| 6 | 14 | 2039.1 | 1307.5 | 0.074 | 1958.4 | 1141.7 |
| 7 | 40 | 2680.9 | 968.6 | 0.118 | 1827.5 | 908.7 |
| 8 | 13 | 2329.3 | 1617.7 | 0.040 | 2099.2 | 1495.9 |
| 9 | 15 | 787.3 | 170.6 | 0.094 | 871.9 | 147.8 |
| 10 | 7 | 642.2 | 1276.7 | 0.240 | 1100.1 | 1179.9 |
| 11 | 10 | 1385.7 | 993.9 | 0.073 | 1508.0 | 891.1 |
| 12 | 11 | 391.4 | 235.3 | 0.101 | 450.6 | 262.8 |

Table 2. Capital Values using the Exp4 and Normal distributions

Given the low number of data points, it was anticipated that goodness of fit would, in general, be poor. In reality, a more optimistic result emerged. The significance levels for a two-tailed test were all below the 10% level (i.e. 5% in each tail), and many of them were within the 5% limit (2.5% in each tail). These results indicate that The *Exp4* distribution is, indeed, a good fit to the data. There is one exception: data set 10. This has only 7 aggregated losses, and the *Exp4* distribution is biased towards the lower values. In the *LDA* random sampling, very high values still figure significantly, and these inflate the capital value. Clearly *Exp4* is inadequate for this particular data set. Figure 5 shows the densities for best and worst *Exp4* fits (left-hand is best, right-hand is worst). The blue profiles are the *Exp4* densities and the red profiles are empirical densities.

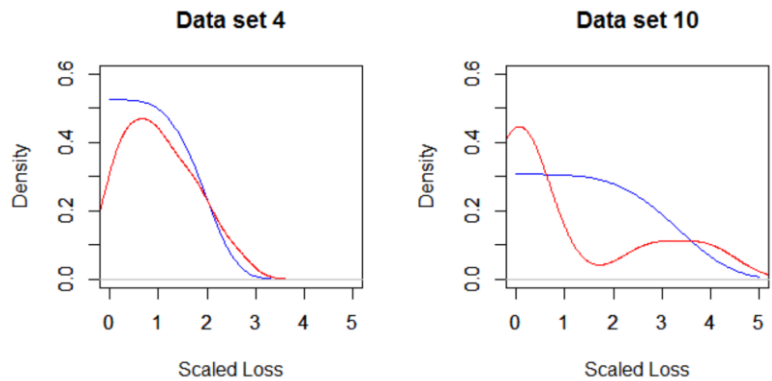

Figure 5. Comparison of *Exp4* and Normal densities for best and worst fits:
(LHS=Best, RHS=Worst; blue = *Exp4*, red = empirical)

In Figure 5, the 'worst' fit fails so badly because the empirical distribution is bimodal. A possible better fit is to use an *Exp4* mixture with density of the form $(1-p) \times Exp4(s_1) + p \times Exp4(s_2)$, where $p$ is a real-valued weight in the range (0,1), and $s_1$ and $s_2$ are the parameters of two *Exp4* distributions. The idea of a mixture distribution was used in [15], but with Normal distributions. In that case each loss was replaced by random samples from Normal distributions. The results were only just acceptable, and replacing the Normal distribution by *Exp4* distributions is a possible way forward. The 'best' fit works well because the bulk of the probability mass falls in the part of the domain of the *Exp4* distribution that can generate the highest probability in random sampling. The 'best' fit is a good illustration of the main criteria for a distribution to work well with aggregated losses: a high probability of generating lower valued losses with a low probability of generating higher valued losses.

The extended distribution *ExpN* with $n \to \infty$ is an acceptable fit in most cases. A general trend with the *ExpN* distribution is for the capital values to decrease and the *TN-A* measures to increase as $n$ increases, but both at a decreasing rate of change. If $n > 12$ the gain in lower capital becomes comparable with the stochastic error inherent in the *LDA* process.

All capitals *ExpN* in Table 2 are less than their corresponding *Exp4* capitals except for Data Set 12. It is likely that sufficient medium-sized losses are sampled in the *ExpN* case ($n \to \infty$) to inflate the *ExpN* capital. The mean reduction, excluding Data Set 12, is 8.1%.

## 5. DISCUSSION

This research was prompted by the unsuitability of 'traditional' "fat-tailed" distributions for modelling conduct risk losses. Regulations [1, 2] require operational risk losses to be aggregated by root event, and for conduct risk losses the result is a very small number of huge losses. Effectively the loss distribution has lost its body and is left with only a tail. The resulting capital values obtained using the LDA process (*Algorithm LDA*, section 3.2) are always much greater than is merited by anticipated future losses.

A suggested solution is to use exactly the extreme opposite of a "fat-tailed" distribution, namely a "very-thin-tailed" distribution. Such a class of distribution decays faster than the Normal distribution for large losses. *Exp4* is a simple extension of the Normal density, and symbolic computation is useful in generating the required density, and integrating it to derive its distribution. If the domain is restricted to positive real numbers only, *Exp4* satisfies the "low loss = high probability; high loss = low probability" criterion stated in section 3.4.

The numerical results are encouraging. They are consistent with expectations based on the previous results which did not involve aggregated losses. Furthermore they are consistent with projected conduct risk losses for the following year.

Using the proposed *ExpN* distribution ($n > 4$) is not recommended. It must be noted that a density with an $exp(-x^n)$ term when $n$ is large (and in this context $n > 4$ is large) appears to be very contrived! The particular case $exp(-x^{100})$ illustrates the point well. It represents extreme overfitting. It is better, in principle, to use a simpler distribution and accept a higher capital value as a more prudent guard against future losses. If the Regulator thinks that capital is too low due to a contrived calculation method, an 'add-on' can be imposed, and the effort in finding a distribution that results in the lowest possible capital is wasted. Therefore the *Exp4* distribution suffices; it works well enough.

## APPENDIX A

Mathematica scripts to calculate the *Exp4* density

```
f1[x_, s_] := E^(-(x^4/(2*s^4)))
k1 = Integrate[f1[x, s], {x, 0, Infinity}, Assumptions -> {Re[s^4] > 0}];
k = FullSimplify[k1, Assumptions→{Im[s]==0, Re[s]>0}]
```

$2^{1/4}\, s\, \text{Gamma}\left[\dfrac{5}{4}\right]$

```
f[x_, s_] := f1[x, s] / (2^(1/4) s Gamma[5/4])

Plot[f[x, 0.2], {x, 0, 0.5}]
```

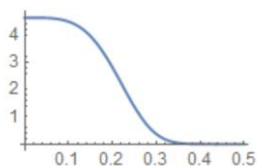

Mathematica scripts to calculate the *Exp4* distribution

```
F[x_, s_] := FullSimplify[Integrate[f[t, s], {t, 0, x}],
    Assumptions → {Im[s] == 0, Re[s] > 0, Im[x] == 0, Re[x] > 0}]

F[x, s]
```

$1 - \dfrac{\text{Gamma}\left[\dfrac{1}{4},\, \dfrac{x^4}{2 s^4}\right]}{\text{Gamma}\left[\dfrac{1}{4}\right]}$

```
Plot[F[x, 0.2], {x, 0, 0.5}]
```

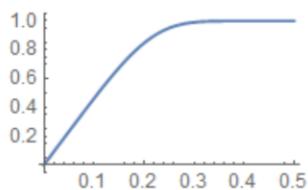

## APPENDIX B

R functions to implement the *Exp4* probability distribution.

Functions *dNormExp4()* and *pNormExp4()* correspond to Equations (2) and (3) respectively. Function *qNormExp4()* is the inverse of *pNormExp4()* and uses a search method to do the inversion. This makes it slow, and a direct method would be preferable. Function *rNormExp4()* generates *Exp4*-distributed random numbers by inverting the distribution function.

```
dNormExp4 <- function(x,s)
{ (1/(s*2^0.25*gamma(5/4)))*exp(-0.5*(x/s)^4)}

pNormExp4 <- function(x,s)
{
  p <- 1 - (1/(gamma(1/4)))*gsl::gamma_inc(1/4, x^4/(2*s^4))
   return(p)
}

qNormExp4 = function(p, s, eps = 1e-10)
{
  lim <- 20*s
  x <- rootSolve::uniroot.all(function(z) { pNormExp4(z, s) - p }, interval = c(eps, lim - eps) )
  if (length(x)>1)   {return(x[1])}
  else   {return(x)}
}

rNormExp4 = function(n, s)
{
  nums <- runif(n)
  rn <- unlist(lapply(1:n, function(z_) {qNormExp4(nums[z_], s)}))
   return(rn)
}
```

## APPENDIX C

Formal hypothesis test for use with the *TN-A* test.

Let $X = \{x_1, x_2, \ldots, x_n\}$ be a sample of size *n* drawn from a random variable *V* that has some probability distribution $\Delta$ (it is implied that this distribution has well-defined density and distribution functions *f* and *F* respectively. Formulate null and alternative hypotheses for a 2-tailed test as follows.

Null hypothesis ($H_0$):         $V \sim \Delta$  (i.e. *V* has a $\Delta$-distribution)
Alternative hypothesis ($H_1$):  $V !\sim \Delta$ (i.e. *V* has any other distribution)

Given a calculated value, *t*, of the *TN-A* statistic and a *p*% critical value, $t_p$, of the *TN-A* statistic:

Reject $H_0$ at *p*% if $t > t_p$
Reject $H_1$ at *p*% if $t \leq t_p$.

Some useful 2-tail *TN-A* critical values are: 0.014 (1%), 0.068 (5%) and 0.131 (10%). Others may be found by inverting equation (5).